\documentclass[conference]{IEEEtran}
\IEEEoverridecommandlockouts
\usepackage[utf8]{inputenc}
\usepackage{hyperref}
\usepackage{xspace}
\usepackage{booktabs}
\usepackage{multirow}
\usepackage[export]{adjustbox}
\usepackage{array}
\usepackage{subfig}

\graphicspath{{figure/}}

\newcommand{\eg}{{e.g.,}\xspace}

\newtheorem{theorem}{$\blacksquare$ \bf{Finding}}  
\newtheorem{lemma}{$\blacksquare$ \bf{Guideline}}
\newtheorem{tip}{$\blacksquare$ \bf{Lesson}}

\usepackage{amsmath,amsfonts,bm}

\def\eqref#1{equation~\ref{#1}}

\def\1{\bm{1}}

\DeclareMathAlphabet{\mathsfit}{\encodingdefault}{\sfdefault}{m}{sl}
\SetMathAlphabet{\mathsfit}{bold}{\encodingdefault}{\sfdefault}{bx}{n}

\usepackage{cite,balance}
\usepackage{url}
\usepackage{hyperref}
\usepackage{amsmath,amssymb,amsfonts}
\usepackage{algorithmic}
\usepackage{graphicx}
\usepackage{textcomp}
\usepackage{xcolor}
\usepackage{booktabs}
\usepackage{graphicx}
\usepackage{caption} 
\usepackage[ruled,vlined]{algorithm2e}
\SetKwInput{Input}{Input}
\SetKwProg{kwSystem}{System executes}{:}{}
\SetKwProg{kwClient}{ClientUpdate}{:}{}
\SetKwProg{kwServer}{ServerAggregation}{:}{}

\def\M{\mathcal{M}}
\def\de{\overset{\Delta}{=}}

\def\BibTeX{{\rm B\kern-.05em{\sc i\kern-.025em b}\kern-.08em
    T\kern-.1667em\lower.7ex\hbox{E}\kern-.125emX}}
\begin{document}

\title{Personalized Federated Recommender Systems with Private and Partially Federated AutoEncoders\\

\thanks{Qi Le, Jie Ding, and Vahid Tarokh were supported in part by the Office of Naval Research under grant number N00014-21-1-2590.}
}

\author{\IEEEauthorblockN{Qi Le}
\IEEEauthorblockA{\textit{College of Science and Engineering} \\
\textit{University of Minnesota-Twin Cities}\\
Minneapolis, USA \\
le000288@umn.edu}
\and
\IEEEauthorblockN{Enmao Diao}
\IEEEauthorblockA{\textit{Electrical and Computer Engineering} \\
\textit{Duke University}\\
Durhm, USA \\
enmao.diao@duke.edu}
\and
\IEEEauthorblockN{Xinran Wang}
\IEEEauthorblockA{\textit{College of Science and Engineering} \\
\textit{University of Minnesota-Twin Cities}\\
Minneapolis, USA \\
wang8740@umn.edu}
\and
\IEEEauthorblockN{Ali Anwar}
\IEEEauthorblockA{\textit{College of Science and Engineering} \\
\textit{University of Minnesota-Twin Cities}\\
Minneapolis, USA \\
aanwar@umn.edu}
\and
\IEEEauthorblockN{Vahid Tarokh}
\IEEEauthorblockA{\textit{Electrical and Computer Engineering} \\
\textit{Duke University}\\
Durhm, USA \\
vahid.tarokh@duke.edu}
\and
\IEEEauthorblockN{Jie Ding}
\IEEEauthorblockA{\textit{School of Statistics} \\
\textit{University of Minnesota-Twin Cities}\\
Minneapolis, USA \\
dingj@umn.edu}
}

\maketitle

\begin{abstract}

Recommender Systems (RSs) have become increasingly important in many application domains, such as digital marketing. Conventional RSs often need to collect users' data, centralize them on the server-side, and form a global model to generate reliable recommendations. However, they suffer from two critical limitations: the personalization problem that the RSs trained traditionally may not be customized for individual users, and the privacy problem that directly sharing user data is not encouraged. We propose Personalized Federated Recommender Systems (PersonalFR), which introduces a personalized autoencoder-based recommendation model with Federated Learning (FL) to address these challenges. PersonalFR guarantees that each user can learn a personal model from the local dataset and other participating users' data without sharing local data, data embeddings, or models. PersonalFR consists of three main components, including AutoEncoder-based RSs (ARSs) that learn the user-item interactions, Partially Federated Learning (PFL) that updates the encoder locally and aggregates the decoder on the server-side, and Partial Compression (PC) that only computes and transmits active model parameters. Extensive experiments on two real-world datasets demonstrate that PersonalFR can achieve private and personalized performance comparable to that trained by centralizing all users' data. Moreover, PersonalFR requires significantly less computation and communication overhead than standard FL baselines.
\end{abstract}


\section{Introduction}

As a result of the fast rise and widespread use of internet services and applications, Recommender Systems (RSs) have become essential in many fields, including digital marketing, customized health, and data mining~\cite{adomavicius2005toward}. RSs can assist users in making efficient use of available information. Most recent works on RSs require all the data from multiple domains to be shared and the calculation for model training to be performed centrally. However, this centralization has two significant limitations: 1) The individual users may not receive personalized models since the global model that the RSs learned traditionally needs to account for the data heterogeneity among users~\cite{li2020federated}. 2) As users' data held on the server may be inadvertently disclosed or exploited, the centralized training naturally leads to privacy concerns~\cite{zhang2014privacy}. 

Federated learning (FL)~\cite{konevcny2016federated,mcmahan2017communication,HeteroFL,diao2021semifl} is a distributed machine learning framework that allows users to train models without direct data sharing. By distributing the model training process to local clients, FL utilizes local compute resources and ensures that user data remain on the client's devices. Federated averaging (FedAvg)~\cite{mcmahan2017communication} is a popular training algorithm that allows multiple local updates, which may facilitate convergences and communication efficiency. However, due to potential data distributional heterogeneity~\cite{karimireddy2020scaffold, konevcny2016federated}, FL may need better convergence~\cite{zhao2018federated} and to provide personal recommendations for different users.



Motivated by the aforementioned issues, this work proposes Personalized Federated Recommender Systems (PersonalFR), which introduces a personalized AutoEncoder-based recommendation model with Federated Learning (FL). The basic architecture of PersonalFR is training the personalized encoder for each client to establish client-specific mapping information and averaging the decoder on the server side. We aim to use local computing resources, complete local training without data sharing, and provide heterogeneous users with personalized recommendations.
Specifically, the server picks a group of accessible clients and provides them with the global decoder. Then, each client updates the local AE model with respect to the local objective function for several epochs. After that, the server leverages Partially Federated Learning (PFL) and Partial Compression (PC) to aggregate local decoders and update the global decoder model. Once the global decoder converges reasonably well, each client uses its customized AE model to obtain recommendations. The suggested PersonalFR with PFL shows faster convergence than standard FL baselines in our extensive experiments. Moreover, using the PC component, PersonalFR requires substantially less processing and transmission overhead than conventional FL baselines. Our contributions are summarized below.

\begin{itemize}
\item We present a new framework of Recommender Systems (RSs), which can provide precise, private, and personalized recommendations without sharing local data, data embeddings, or models.
\item By leveraging Partially Federated Learning (PFL), we show that our PersonalFR can outperform FedAvg and achieve private and personalized performance comparable to that trained by centralizing all the users’ data.
\item We propose Partial Compression (PC) that only computes and transfers the active model parameters. Our experimental results show that the PersonalFR compresses the computation overhead around $1.25 \times$ to $1.9 \times$ over FedAvg and communication overhead around $2.5 \times$ to $27 \times$ over FedAvg.
\end{itemize}

\vspace{-4pt}
\section{Related Work}
\vspace{-2pt}
\subsection{Recommender Systems}
Recommendation Systems (RSs) may be divided into three rough categories~\cite{jannach2010recommender}: content-based filtering, collaborative filtering, and hybrid methods. Content-based filtering~\cite{rocca2019introduction} predicts users' preferences based on their side information, \eg personal information. Collaborative filtering~\cite{rocca2019introduction} leverages user-item interactions and the items that users with comparable preferences favored to infer users' preferences for particular items. The hybrid methods~\cite{diao2021privacy} combine content-based filtering and collaborative filtering. Our proposed method is a collaborative filtering recommender system that uses user-item interactions.

\subsection{Personalized Federated Learning}

Personalized Federated Learning has been a solution to modeling heterogeneous data to provide users with personalized models.
There are two rough categories of personalized federated learning~\cite{chen2022self}: global model personalization and personalized models. The first category trains a good globally-shared FL model, and then the trained global FL model is adapted locally for each FL client~\cite{mansour2020three}.
For example, FAug~\cite{jeong2018communication} tries to mitigate statistical heterogeneity across clients' datasets. MAML~\cite{fallah2020personalized} seeks to develop a globally generalizable model. The second category focuses on training personalized FL models for clients. For example, FedMD~\cite{li2019fedmd} and SPIDER~\cite{mushtaq2021spider} aim to establish client-specific model architectures via knowledge distillation and neural architectural search, respectively. FedAMP~\cite{huang2021personalized} utilizes similarity between clients' data distributions to enhance the performance of tailored models. Self-FL~\cite{chen2022self} automatically tunes clients' local model initialization, training steps, and server aggregation based on balancing the inter-client and intra-client model uncertainty from a Bayesian hierarchical modeling perspective. We refer to~\cite{chen2022self} for a more detailed literature review on personalized federated learning.
The existing federated personalization methods often require additional resources in computation and communication. This is particularly a concern for many practical applications of recommender systems that require a large model to address a large scale of users, items, and ratings heterogeneously distributed among users~\cite{jannach2010recommender}. It has motivated our work to study a more efficient personalized federated learning solution to recommender systems.

\vspace{-10pt}
\begin{figure}[htbp]
\centerline{\includegraphics{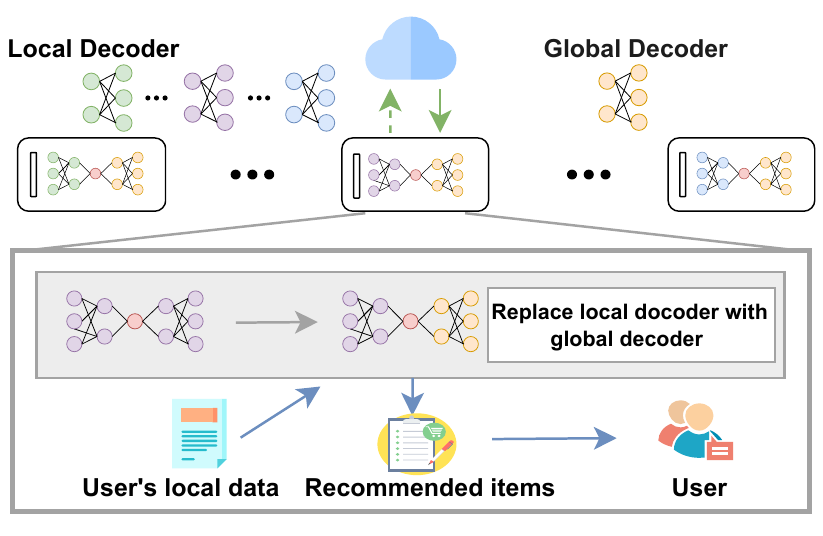}}
\caption{Personalized Federated Recommender Systems. User can obtain improved personal recommendations by leveraging the data from global domain without sharing the local data and encoder part of the model.}
\label{schema}
\end{figure}

\begin{figure*}[htbp]
\centerline{
\includegraphics[scale=0.75]{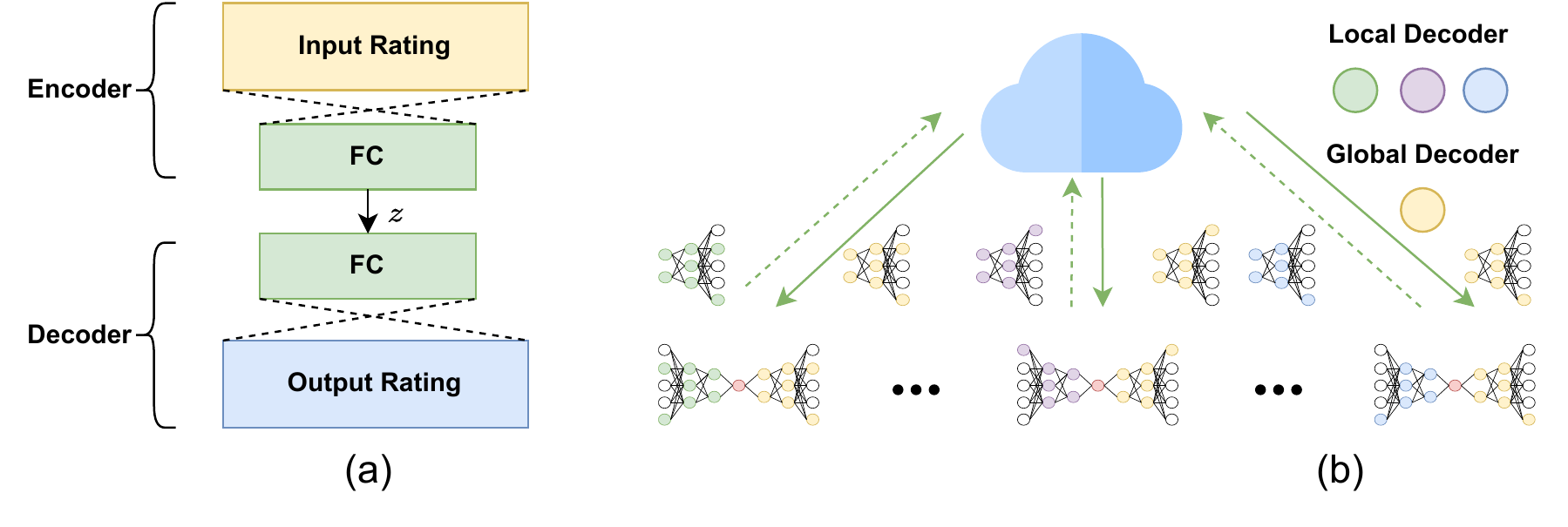}}
\vspace{-14pt}
\caption{(a) Illustration of AutoEncoder. The input rating dimension is the number of items of that domain, while the output dimension is the same as the input dimension. (b) Illustration of Partially Federated Learning and Partial Compression. Only the active parameters, as shown in Expression~(\ref{active_parameters}), will be transmitted.}
\label{method_figure}
\end{figure*}

\setlength{\abovecaptionskip}{0.cm}

\vspace{-16pt}
\section{Method}
In this section, we present the framework architecture for PersonalFR that offers personalized recommendations shown in Figure~\ref{schema}.

\vspace{-4pt}
\subsection{Problem Formulation}
\vspace{-4pt}
Our PersonalFR is a rating-based collaborative filtering recommender system that predicts users’ explicit preference ratings for items based on user-item interactions. We define $\textit{U}$ $\de$ \{$u_1$ , \dots, $u_k$\} be the set of $k$ users and $V$ $\de$ \{$v_1$, \dots, $\textit{$v_n$}$\} be the set of $n$ items. Then, we have a user-item interaction matrix $R=[r_{i, j}]_{1 \leq i \leq k, 1 \leq j \leq n} \in \mathbb{R}^{k \times n}$, where each element $r_{i, j}$ indicates the rating score that user $u_i$ assigns to the item $v_j$. Utilizing a recommender system, we can predict $\hat{r}_{i, j}$ for user $u_i$ attributing to the item $v_j$. The objective of training a recommender system is to minimize the average of following training loss for all the rating scores
\begin{equation}
\mathop{}\sum\limits_{1 \leq i \leq k, 1 \leq j \leq n, r_{i, j} \neq 0} \ell(\hat{r}_{i, j}, r_{i, j}) \label{eq}
\end{equation}
over model parameters that define $\hat{r}_{i, j}$'s, where $\ell(\cdot)$ is the loss function. When calculating the loss value, our PersonalFR will mask out the unrated user-item interactions and minimize the average of the loss values between the rated user-item interactions and the fitted user-item interactions. We will use quadratic loss for regression and cross-entropy loss for classification in the experimental study.

\vspace{-4pt}
\subsection{AutoEncoder}
\vspace{-4pt}
AutoEncoder (AE) has many successful applications in RSs~\cite{sedhain2015autorec, diao2021privacy}.
AE encodes a high-dimensional input signal into a low-dimensional hidden representation, and then decodes it into an output representation. The structure of AE is shown in Figure~\ref{method_figure}. In addition, AE considers rating matrices as tabular data, where rows represent subjects and columns represent features. We define $\mathcal{X} \de \{x_1, \dots, x_k\}$ as the set of all user rating vectors, where $x_i \de (r_{i,1} , \dots, r_{i,n}) \in \mathbb{R}^{n}$ denotes the rating scores that user $u_i$ assigns to the items $v_1 \dots v_n$. All user rating vectors are sparse vectors, where the unrated user-item interactions are zeros. We define $\hat{\mathcal{X}} \de \{\hat{x}_1, \dots, \hat{x}_k\}$ as the set of all predicted user rating vectors, where $\hat{x}_i \de (\hat{r}_{i,1} , \dots, \hat{r}_{i,n}) \in \mathbb{R}^{n}$ represents the predicted rating scores that user $u_i$ gives to the items $v_1 \dots v_n$. Our AE takes the vector $x_i$ as the input and produces the vector $\hat{x}_i$ as the output. The output vector has the same dimension as the input vector. In particular, AE consists of an encoder $z$ = $E(x_i):R^{n} \rightarrow R^{d_{\textrm{hidden}}}$ and a decoder $\hat{x}_i$ = $D(z):R^{d_{\textrm{hidden}}} \rightarrow R^{n}$, where $n$ and $d_{\textrm{hidden}}$ represent the dimensions of the input/output vector and the latent vector $z$, respectively. 

\vspace{-6pt}
\subsection{Personalized Federated Recommender Systems (PersonalFR)}
\vspace{-2pt}
\begin{algorithm}[htbp]
\small
\SetAlgoLined
\DontPrintSemicolon
\Input{Data $\mathcal{X}$ distributed on $M$ local clients $(\mathcal{X} = \mathcal{X}_{1} \cup \dots \cup \mathcal{X}_{m})$, fraction $C$ of selected clients per communication round, local minibatch size $B$, learning rate $\eta$, total number of communication rounds $T$, number of local epochs $K$, local encoders distributed on $M$ clients parameterized by $E_m$ ($m=1,\ldots,M$), globally shared decoder parameterized by $D_g$.}
\kwSystem{}{
Initialize $D_g^0$\;
\For{\textup{each client }$m \in \M$\textup{ \textbf{in parallel}}}{
Initialize $E_{m}^{0}$\;
Send indices of rated user-item interactions $I_m$ to the server
}
Server records all clients' indices $I$ = $\{I_1, \dots, I_m\}$\;
\For{\textup{each communication round }$t = 1, \dots, T$}{
$\M^t \leftarrow \max(C \cdot M, 1)$ clients sampled from $\M$\;
Initialize decoder parameters set $\mathcal{S}^t \leftarrow$  $\varnothing$ \;
\For{\textup{each client }$m \in \M^t$\textup{ \textbf{in parallel}}}{
$E_{m}^{t} \leftarrow E_{m}^{t-1}$\; 
$D_{m, \text{active}}^{t} \leftarrow$ Server sends active parameters based on $D_{g}^{t-1}$ and $I_m$(See Expression~(\ref{active_parameters}))\;
$E_{m}^t$, $D_{m, \text{active}}^{t}$ $\leftarrow$ ClientUpdate($E^{t}_{m}, D_{m, \text{active}}^{t}$)\;
$\mathcal{S}^{t}$ $\leftarrow$ $\mathcal{S}^{t} \cup \{D_{m, \text{active}}^{t}\}$ \;
}
$D^{t}_{g} \leftarrow$ ServerAggregation($D^{t-1}_{g}$, $\mathcal{S}^{t}$)\;
}
}

\kwClient{$(E^{t}_{m}, D_{m, \textup{active}}^{t})$}{
$B_m$  $\leftarrow$  Split  local  data $\mathcal{X}_{m}$  into  batches  of  size  $B$\;
\For{\textup{each local epoch }$l = 1, \dots, K$}{
\For{\textup{batch} $b_{m} \in B_m$}{
$E_{m}^t \leftarrow E_{m}^t - \eta \nabla_{E} L( D_{m, \text{active}}^{t}(E_{m}^t), b_{m})$\;
$D_{m, \text{active}}^{t} \leftarrow D_{m, \text{active}}^{t} - \eta \nabla_{D} L(D_{m, \text{active}}^{t}(E_{m}^t), b_{m})$\;
($L$: total loss for $b_m$ based on Formula~(\ref{eq}))
}
}
Return $E_{m}^t$, $D_{m, \text{active}}^{t}$
}
\kwServer{($D^{t-1}_{g}$, $\mathcal{S}^{t}$)}{
$D^{t}_{g} \leftarrow D^{t-1}_g$\;
\For{\textup{each active parameters} $D^{t}_{m, \textup{active}} \in \mathcal{S}^{t}$}{
$D^{t}_{m} \leftarrow$ Use $D^{t}_{m, \text{active}}$ and $D^{t-1}_{m}$ to update $D^{t}_{m} $ according to Equation~(\ref{refill_active_parameters})\;
$D_{g}^{t} \leftarrow D_{g}^{t}$ + $\frac{1}{\left| \mathcal{S}^{t} \right|} D^{t}_{m}$ ($\left| \mathcal{S}^{t} \right|$: cardinality of $\mathcal{S}^{t}$)\; 

}
Return $D_{g}^{t}$
}

\caption{PersonalFR: Personalized Federated Recommender Systems}
\label{alg:dynamicfl}
\end{algorithm}


In this section, two critical components of PersonalFR are described. The first key component is the personalized partial update of the client and server model, which we refer to as Partially Federated Learning (PFL). The second key component is the optimized computation and communication during the training process, which we call Partial Compression (PC). PC is built upon the PFL, and these elements contribute together to make PersonalFR perform better and need fewer resources than FedAvg for recommender systems. Furthermore, we summarize the pseudocode of the PersonalFR in Algorithm 1 and present the workflow of PersonalFR at the end of this section. 

Before describing the details of two key components, PFL, and PC, we first introduce the general settings and notations. Let $\M$ be the set of the clients, whose cardinality is $M$. Each client owns a unique encoder and shares a global decoder. For a generic decoder, we define the vector $H \de (h_1 , \dots, h_q) \in \mathbb{R}^{q}$ as the output of the last hidden layer, $\mathcal{W} \de \{W_1, \dots, W_p\}$ as the full set of weights, $\mathcal{B} \de \{B_1, \dots, B_p\}$ as the full set of biases, where $p$ is the number of layers. Each element of $\mathcal{W}$ is a matrix, and each element of the $\mathcal{B}$ is a vector. In the sequel, for the above quantities associated with a particular client $m$, we will put a subscript $m$ to highlight such association.



Let $\M$ be the set of the clients, whose cardinality is $M$. For client $m \in \M$, we use $\mathcal{X}_{m}$ and $\hat{\mathcal{X}}_{m}$ to represent all the user rating vectors and all the predicted user rating vectors within the client $m$, respectively, such that $\mathcal{X}$ = $\cup_{m \in \M}\mathcal{X}_m$ and $\hat{\mathcal{X}}$ = $\cup_{m \in \M}\hat{\mathcal{X}}_m$. For client $m$, the output vector $\hat{x}_{m, i}$ = $(\hat{r}_{i,1} , \dots, \hat{r}_{i,n}) \in \mathbb{R}^{n}$ from the AE model is generated from
\begin{align}
    \hat{x}_{m, i} = W_{m, p} \cdot H_m + B_{m, p},
\end{align}
where $W_{m, p} \in \mathbb{R}^{n \times q}$ and $B_{m, p} \in \mathbb{R}^n$ represent the weight matrix and the bias of the output layer associated with the client $m$, respectively.



As shown in Table~\ref{dataset_details}, the observable data associated with a typical recommender system is highly sparse. As a result, for every client, only a tiny fraction of items, say $\{v_1, \dots, v_{n^{\prime}}\}$, where $n^{\prime} \ll n$, are rated by at least one user of that client. Thus, only the submatrix $W^{\prime}_{m, p} \in \mathbb{R}^{n^{\prime} \times q}$ of the weight matrix $W_{m, p}$ and the subvector $B^{\prime}_{m, p} \in \mathbb{R}^{n^{\prime}}$ of the bias vector $B_{m, p}$ connected to the rated items will be effectively used to predict rating scores. In line with that, only $W^{\prime}_{m, p}$ and $B^{\prime}_{m, p}$ will be updated in the back-propagation during the local training.

Then, we elaborate on more details of PFL, which trains a personalized encoder for each client to generate client-specific encoder mapping and averages the decoder on the server side to gain model improvement from other clients' models. More specifically, for each client, PFL trains its personalized encoder together with the global decoder on the local dataset. Then, only the decoder will be processed during the transmission and server aggregation. This is different from FedAvg that transfers and averages the whole model. The unshared personalized encoder of each client can extract the unique features of each client's input. Consequently, the procedure for calculating predicted rating scores differs for FedAvg and PersonalFR. In particular, for client $m$ in FedAvg, the predicted rating scores are $\hat{x}_{m, i}$ = ${D(E(x_{i}))} \in \mathbb{R}^n$ for user $u_i$ attributing to all $n$ items, where $D(\cdot)$ is the global decoder and $E(\cdot)$ is the global encoder. In contrast, in PersonalFR, the predicted rating scores are $\hat{x}_{m, i}$ = ${D(E_{m}(x_{i}))} \in \mathbb{R}^n$ for user $u_i$ attributing to the all $n$ items, where $D(\cdot)$ is the global decoder and $E_{m}(\cdot)$ is the personalized encoder of the client containing user $u_i$. Moreover, it is worth mentioning that keeping the private personalized encoder for all clients not only helps us improve the local model's performance but also helps us with some security parts. For example, it would be challenging to recover the user data by possible attackers knowing the parameters of the globally shared decoder and the output vector for a client. 



Next, we show how PC is built upon the PFL to further reduce computation and communication costs. In PersonalFR, we only update and transfer the active parameters $D_{\text{active}}$ of a generic decoder within the training process. The active parameters $D_{m, \text{active}}$ for the client $m$ can be expressed as
\begin{align}
D_{m, \text{active}} \de \{W_{m, 1}, B_{m, 1}, \dots, W^{\prime}_{m, p}, B^{\prime}_{m, p}\},
\label{active_parameters}
\end{align}
which represents all the weight matrices and bias vectors that will be updated during the local back-propagation of the decoder $D_m$.
Here, $W_{m,p}$ and $B_{m,p}$ in $D_{m, \text{active}}$ are the weight matrix and bias vector of the output layer, respectively. On the other hand, in FedAvg, all the parameters of the decoder of client $m$ need to be updated and transferred during the training stage, which can be represented by
\begin{align}
D_{m} \de \{W_{m, 1}, B_{m, 1}, \dots, W_{m, p}, B_{m, p}\}
\end{align}
$D_{m, \text{active}}$ only occupies a tiny portion, e.g., as small as 3.7\% (in our experiment study), of $D_m$. This is because the weight matrix $W_{m, p}$ and bias vector $B_{m, p}$ of the output layer occupy a significant portion of decoder parameters. By shrinking them to $W^{\prime}_{m, p}$ and $B^{\prime}_{m, p}$ using PC, we can greatly reduce computing and communication resources. Besides, during the server aggregation step at communication round $t$, we need first to update $W^{t}_{p}$ and $B^{t}_{p}$ of the output layer of $D^{t}_{m}$ for each participating client. More particularly, for client $m$, we update $W^{t}_{m, p}$ and $B^{t}_{m, p}$ using $W^{\prime t}_{m, p}$ and $B^{\prime t}_{m, p}$, respectively. The following equation illustrates the procedure
\begin{align}
W^{t}_{m, p} = W^{\prime t}_{m, p} \cup (W^{t-1}_{m, p} \backslash W^{\prime t-1}_{m, p})  \nonumber\\
B^{t}_{m, p} = B^{\prime t}_{m, p} \cup (B^{t-1}_{m, p} \backslash B^{\prime t-1}_{m, p}) ,
\label{refill_active_parameters}
\end{align}
where $W^{t-1}_{m, p}$ represents the weight matrix of the output layer of $D^{t-1}_{m}$ at communication round $t-1$, $W^{\prime t}_{m, p}$ represents the submatrix of $W^{t}_{m, p}$ connected to the rated items of client $m$ at communication round $t$, and $W^{t-1}_{m, p} \backslash W^{\prime t}_{m, p}$ represents the set of parameters in $W^{t-1}_{m, p}$ but not in $W^{\prime t}_{m, p}$.





At last, we summarize the execution flow of our PersonalFR system. Algorithm 1 presents the pseudocode of the PersonalFR. In the beginning, we initialize the AE models for all clients. Then, to find the active parameters, each client needs to send the indices of the rated user-item interactions to the server, and the server records all clients' indices. Each client can now locate its active parameters and begin using the PC. Afterward, the server selects a batch of available clients and sends the current global decoder. Throughout several local epochs, each client trains the local AE model with respect to the local objective function and uploads the new local decoder parameters using PC to the server. Last, the server updates the global decoder model by utilizing the selected clients' active parameters and the previously recorded indices. The above training procedure is repeated until the global decoder converges. Finally, in the prediction stage, each client utilizes its personalized AE model to obtain recommendations.

\vspace{-2pt}
\section{Experiments}
\subsection{Experimental Setup}
\textbf{Models and Datasets}. We conduct our experiments on two public datasets: MovieLens1M (ML1M)~\cite{harper2015movielens}, which is a dataset of movie ratings, and Anime~\cite{Anime_kaggle}, which is a dataset of anime ratings. The detailed attributes of these two datasets are listed in Table~\ref{dataset_details}. We filter out users and items with fewer than $20$ ratings for the Anime dataset and pick the first $6000$ users. The details of hyperparameters for model training are listed in Table~\ref{hyperparameters_table}. We have the following control settings.

1) Different number of clients. We evaluate the performance of PersonalFR and FedAvg under various numbers of clients and data heterogeneity scenarios. While the numbers of clients are different, the total amounts of the available data remain the same for ML1M and Anime datasets, respectively. As the number of clients increases, each client will own fewer users and less available data.

2) Explicit versus implicit feedback~\cite{hu2008collaborative}. The explicit feedback is the default rating ($1$-$5$ for the ML1M dataset, $1$-$10$ for the Anime dataset). In contrast, the implicit feedback is the binarized rating (positive if greater than $3.5$ for the ML1M dataset, positive if greater than 8 for the Anime dataset). We regard the explicit feedback as the regression task and the implicit feedback as the binary classification task. We use the $l_2$-norm as the loss function and the Root Mean Square Error (RMSE) as the evaluation metric for explicit feedback. We use the cross-entropy as the loss function and the Normalized Discounted Cumulative Gain (NDCG) as the evaluation metric for implicit feedback.

3) With versus without compression. We contrast the computation and communication costs of PersonalFR runs on the ML1M and Anime datasets with those of FedAvg.

4) Ablation studies. For each dataset, we train on $80\%$ of the available data and test on the remaining $20\%$. Four random experiments are conducted to report the standard errors of performance metrics, which are all smaller than $4\times 10^{-3}$.

\textbf{Baseline}
We compare the proposed method with two baselines, ‘Joint’ and ‘FedAvg.’ ‘Joint’ refers to the centralized case where a single entity owns all data. The ‘FedAvg’ denotes the case where the standard FL baseline is applied. Our method aims to outperform the ‘FedAvg’ case and perform competitively with the ‘Joint’ case. 

\begin{table}[htbp]
 \centering
 \begin{tabular}{lcccccl}\toprule
    Dataset         & $k$  & $n$  & sparsity\\\midrule
    ML1M    & 6040 & 3706 & 96\% \\
    Anime & 69600 & 9927 & 99\% \\
 \bottomrule
 \end{tabular}
 \vspace{-8pt}
 \caption{Detailed attributes of the ML1M and Anime datasets. Each
 dataset contains $k$ users and $n$ items. Sparsity means the percentage of unrated user-item interactions over all user-item interactions.}
 \label{dataset_details}
\end{table}

\vspace{-14pt}
\begin{table}[htbp]
\centering
\resizebox{\columnwidth}{!}{
\begin{tabular}{@{}ccccccccc@{}}
\toprule
Model               & \multicolumn{8}{c}{AutoEncoder}                                                                                               \\ \midrule
Hidden size         & \multicolumn{8}{c}{{[}$n$,256,128{]}, {[}128,256,$n${]}}                                                                          \\ \midrule
Global Epoch        & \multicolumn{8}{c}{800}                                                                                                       \\ \midrule
Local Epoch $K$       & \multicolumn{8}{c}{5}                                                                                                         \\ \midrule
Momentum            & \multicolumn{8}{c}{0.9}                                                                                                       \\ \midrule
Weight decay        & \multicolumn{8}{c}{5.00E-04}                                                                                                  \\ \midrule
Number of clients $M$ & \multicolumn{2}{c}{1}        & \multicolumn{2}{c}{100} & \multicolumn{2}{c}{300} & \multicolumn{2}{c}{6040 (1 User / Client)} \\ \midrule
Data                & ML1M          & Anime        & ML1M       & Anime      & ML1M       & Anime      & ML1M         & Anime                       \\ \midrule
Optimizer           & \multicolumn{2}{c}{Adam~\cite{kingma2014adam}}     & \multicolumn{5}{c}{SGD}                                          & \multirow{3}{*}{N/A}        \\ \cmidrule(r){1-8}
Local Batch Size $B$  & 500           & 100          & \multicolumn{5}{c}{10}                                           &                             \\ \cmidrule(r){1-8}
Learning rate       & \multicolumn{2}{c}{1.00E-03} & \multicolumn{5}{c}{1.00E-01}                                     &                             \\ \bottomrule
\end{tabular}}
\centering
\vspace{-8pt}
\caption{Hyperparameters of our experiments for training local models. The size of our encoder is {[}$n$,256,128{]}, where $n$ is the size of the input. The size of our decoder is {[}128,256,$n${]}, where $n$ is the size of the output, the same size as the input.}
\label{hyperparameters_table}
\end{table}

\begin{table*}[htbp]
\centering
\begin{tabular}{@{}cccccc@{}}
\toprule
\multicolumn{2}{c}{Dataset}                & \multicolumn{2}{c}{ML1M}                          & \multicolumn{2}{c}{Anime}                         \\ \midrule
\multicolumn{2}{c}{Metric}                 & RMSE(↓)                 & NDCG(↑)                 & RMSE(↓)                 & NDCG(↑)                 \\ \midrule
Joint                        & Upper Bound & 0.8591(0.0003)          & 0.8466(0.0024)          & 1.1926(0.0007)          & 0.8576(0.0018)          \\ \midrule
\multirow{2}{*}{100 Clients} & FedAvg      & 0.8629(0.0009)          & 0.8514(0.0012)          & 1.2303(0.0014)          & 0.8606(0.0041)          \\
                             & PersonalFR  & \textbf{0.8613(0.0004)} & \textbf{0.8538(0.0018)} & \textbf{1.2121(0.0004)} & \textbf{0.8621(0.0024)} \\ \midrule
\multirow{2}{*}{300 Clients} & FedAvg      & 0.8731(0.0006)          & 0.8486(0.0025)          & 1.2438(0.0012)          & 0.8603(0.0025)          \\
                             & PersonalFR  & \textbf{0.8602(0.0007)} & \textbf{0.8529(0.0017)} & \textbf{1.2247(0.0007)} & \textbf{0.8611(0.0024)} \\ \bottomrule
\end{tabular}
\vspace{-7pt}
\caption{Results of ML1M and Anime datasets for explicit and implicit feedback. ↓ indicates the smaller the better, while ↑ indicates the larger the better.}
\centering
\label{ml1m_anime_res}
\end{table*}
\setlength{\belowcaptionskip}{-0.3cm}
\vspace{-6pt}
\begin{figure*}[htbp]
\centering
{\includegraphics[width=1.0\linewidth]{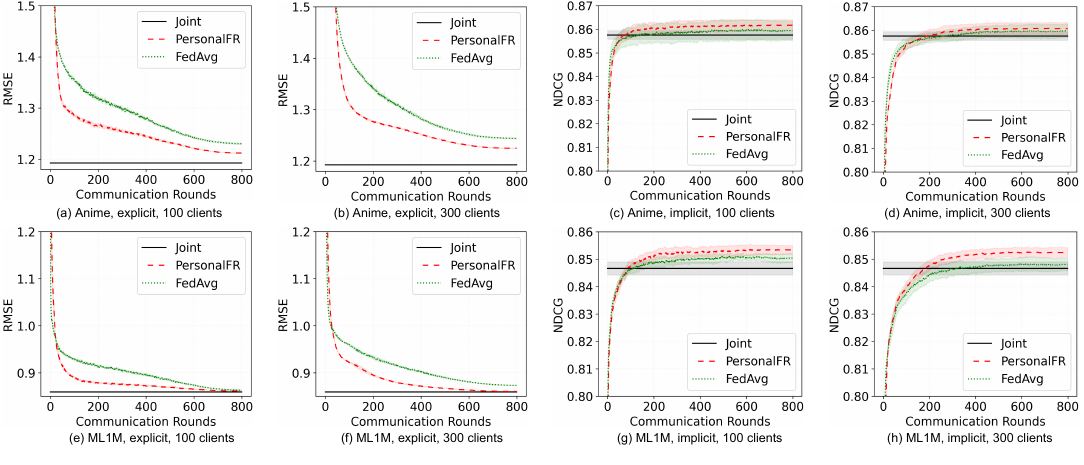}}
\vspace{-18pt}
\caption{Learning curves of the ML1M and Anime datasets for explicit feedback measured with RMSE and implicit feedback measured with NDCG trained by FedAvg and PersonalFR, respectively. (a-d) Anime dataset. (e-h) ML1M dataset.}
\label{ml1m_anime_exp_imp_fig}
\end{figure*}

\subsection{Experimental Results}
\begin{figure}[htbp]
\centering
{\includegraphics[width=1.0\linewidth]{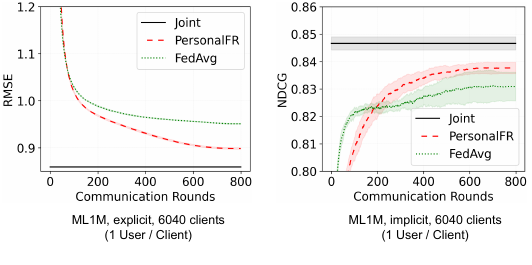}}
\vspace{-20pt}
\caption{Learning curves of the ML1M dataset for explicit (left) and implicit (right) feedback measured with RMSE trained by FedAvg and PersonalFR under 6040 clients(1 User / Client) situation, respectively.}
\label{ml1m_1user_per_client_fig}
\end{figure}

\begin{figure}[htb]
\centering
{\includegraphics[width=1.0\linewidth]{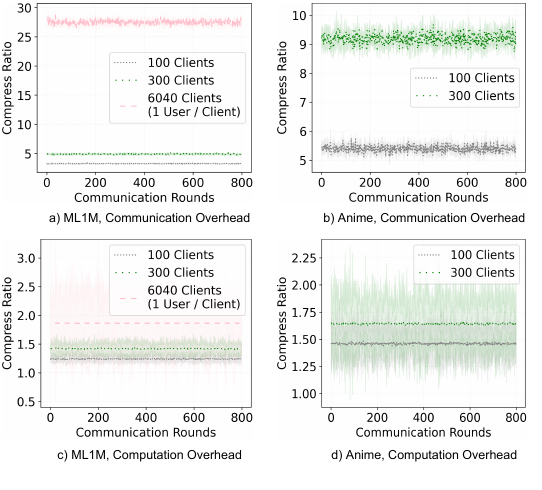}}
\vspace{-20pt}
\caption{The compress ratio of the computation and communication overhead of PersonalFR runs on the ML1M and Anime datasets compared to those of the FedAvg. (a-b) Communication overhead. (c-d) Computation overhead.}
\label{compress_ratio_fig}
\end{figure}

Tables~\ref{ml1m_anime_res} and~\ref{ml1m_1user_per_client_table} lists the experimental outcomes. The standard deviations are shown in brackets with four random experiments. In Figures~\ref{ml1m_anime_exp_imp_fig},~\ref{ml1m_1user_per_client_fig}, and~\ref{compress_ratio_fig}, we depict evaluations across various contexts. We offer thorough explanations below.

\begin{table}[htbp]
\begin{tabular}{@{}cccc@{}}
\toprule
\multicolumn{2}{c}{Dataset}                                                                                    & \multicolumn{2}{c}{ML1M}                          \\ \midrule
\multicolumn{2}{c}{Metric}                                                                                     & RMSE(↓)                 & NDCG(↑)                 \\ \midrule
Joint                                                                                            & Upper Bound & 0.8591(0.0003)          & 0.8466(0.0024)          \\ \midrule
\multirow{2}{*}{\begin{tabular}[c]{@{}c@{}}6040   Clients\\      (1 User / Client)\end{tabular}} & FedAvg      & 0.9508(0.0008)          & 0.8321(0.0054)          \\
                                                                                                 & PersonalFR  & \textbf{0.8983(0.0017)} & \textbf{0.8379(0.0021)} \\ \bottomrule
\end{tabular}
\vspace{-8pt}
\caption{Results of ML1M for explicit and implicit feedback under 6040 clients (1 User/Client) situation }
\label{ml1m_1user_per_client_table}
\end{table}
\setlength{\belowcaptionskip}{-0.5cm}

\textbf{Effect of the number of clients}
\ref{ml1m_anime_res}
As shown in Figure~\ref{ml1m_anime_exp_imp_fig}, we record the learning curves of FedAvg and PersonalFR of the ML1M and Anime datasets for explicit and implicit feedback. As a result, our PersonalFR converges faster and better than FedAvg while keeping a personalized encoder. Moreover, the performance of PersonalFR using explicit and implicit feedback is close to centralized training, as shown in Table~\ref{ml1m_anime_res}. 
If we compare the performance of FedAvg and PersonalFR on $300$ clients with that of $100$ clients, we observe a decline in performance for either explicit or implicit feedback. This decline is perhaps because FedAvg has the known issue of gradient divergence; namely, the directions of the gradient updates generated from each selected client can be significantly different~\cite{zhao2018federated, karimireddy2020scaffold}, especially when clients' data are heterogeneous. As the number of clients increases, data distributional heterogeneity and gradient divergence become increasingly problematic.

To further demonstrate our PersonalFR method, we test the situation where each user obtains a private and personalized model for the ML1M dataset. In the ML1M dataset, there are $6040$ users, so we have $6040$ clients. We summarize the results in Table~\ref{ml1m_1user_per_client_table} and show the learning curve for explicit and implicit feedback in Figure~\ref{ml1m_1user_per_client_fig}. Under this scenario, our PersonalFR performs significantly better than the FedAvg for explicit feedback, indicating that personalized encoder mapping is increasingly helpful for predicting user preferences.
Moreover, in the experiments, the hyperparameters are not tuned for optimal performance in the federated setting. Thus, the experimental results could be potentially improved further.

\textbf{Explicit versus implicit feedback}
As shown in Table~\ref{ml1m_anime_res} and~\ref{ml1m_1user_per_client_table}, our PersonalFR outperforms the FedAvg in most scenarios and achieves competitive results in the ‘Joint’ scenario. Furthermore, we can observe that when the number of clients increases, the performance drops of FedAvg and PersonalFR for explicit feedback are more significant than those of implicit feedback according to Table~\ref{ml1m_anime_res} and Table~\ref{ml1m_1user_per_client_table}. Moreover, for both explicit and implicit feedback, the performance of PersonalFR drops less than that of FedAvg. Especially for explicit feedback, the performance of PersonalFR drops significantly less than that of FedAvg. Therefore, PersonalFR has much more advantages than FedAvg for explicit feedback. 


\textbf{With versus without compression}
We conduct experiments for PC with various numbers of client scenarios. We show the results in Figure~\ref{ml1m_1user_per_client_fig}. Our personalFR can compress the computation overhead around $1.25 \times$ to $1.9 \times$ over FedAvg and communication overhead around $2.5 \times$ to $27 \times$ over FedAvg according to the number of clients. Furthermore, the results show that as the number of clients increases, the local dataset for each client becomes sparser, meaning that fewer rated items are observed, and the proposed PC achieves a more significant reduction in computation and communication.

\section{Conclusion and Future Work}
In this work, we propose Personalized Federated Recommender Systems (PersonalFR), which combines a personalized AutoEncoder-based recommendation model with Federated Learning (FL). We demonstrate that by using Partially Federated Learning (PFL), our PersonalFR can surpass the FedAvg and obtain private and customized performance close to that achieved by centralizing all user data. Furthermore, the PersonalFR requires far less computation and communication overhead than the FedAvg by applying Partial Compression (PC).
An interesting future problem is to address the performance decline that occurs as the number of clients increases. In addition, one may examine the newly developed approach to federated recommender systems in the presence of various adversarial attacks.

\balance
\bibliographystyle{IEEEtran}
\bibliography{asilomar_reference}

\clearpage

\end{document}